\documentclass[%
 aip,
 amsmath,amssymb,
 reprint,%
]{revtex4-1}

\usepackage{graphicx}
\usepackage{dcolumn}
\usepackage{bm}

\usepackage[utf8]{inputenc}
\usepackage[T1]{fontenc}
\usepackage{mathptmx}
\usepackage{upgreek}

\begin{document}

\preprint{AIP/123-QED}

\title[Merging of Soap Bubbles]{Merging of Soap Bubbles and Why Surfactant Matters}

\author{Patricia Pfeiffer}
 \affiliation{Institute for Physics, Otto von Guericke University Magdeburg,  Magdeburg 39106, Germany}
\email{patricia.pfeiffer@ovgu.de}
\author{Qingyun Zeng}%
\affiliation{Division of Physics and Applied Physics, School of Physical and Mathematical Sciences, Nanyang Technological University, 637371 Singapore, Singapore}%

\author{Beng Hau Tan}
\affiliation{Low Energy Electronic Systems, Singapore-MIT Alliance for Research and Technology, Singapore 138602, Singapore}%

\author{Claus-Dieter Ohl}
 \affiliation{Institute for Physics, Otto von Guericke University Magdeburg, Magdeburg 39106, Germany}

\date{\today}

\begin{abstract}
The merging of two soap bubbles is a fundamental fluid mechanical process in foam formation. In the present experimental study the liquid films from two soap bubbles are brought together. Once the liquid layers initially separated by a gas sheet are bridged on a single spot the rapid merging of the two liquid films proceed. Thereby the connecting rim is rapidly accelerated into the separating gas layer. We show that  due to the dimple formation the velocity is not uniform and the high acceleration causes initially a Rayleigh-Taylor instability of the liquid rim. At later times, the rim takes heals into a circular shape. However for sufficient high concentrations of the surfactant the unstable rim pinches off microbubbles resulting in a fractal dendritic structure after coalescence.  
\end{abstract}
\maketitle

\date{\today}

The coalescence of two soap bubbles is a fundamental process in the production of foams and is thus crucial for many industrial processes such as waste water treatment \cite{Ferguson1974} or foam separation \cite{Brown1999}. 
Despite the importance of the soap bubble coalescence process for the growth, structure and microscopic properties of foams very few studies have addressed the fluid mechanics of two merging thin filmed bubbles.

When two soap bubbles approach each other, their films deform slightly by the pressure built up from the entrapped gas. Once  this gas sheet is sufficiently thin attractive van-der-Waals forces create a liquid bridge connecting the two liquid films. This may occur at the closest distance of the deformed soap films. The connecting bridge driven by surface tension quickly spreads radially out thereby merging the two films \cite{Horn2011}. The coalescence of the two bubbles involves the deformation of the liquid films, static surface forces, rheology, and hydrodynamics \cite{Dagastine2006, Manica2007, Manica2008PF}.

After the coalescence of the bubbles, they share a single film that over time drains and eventually ruptures. The drainage and the stability of the foam film are governed by intermolecular forces and surface rheology. Additionally, in the presence of spatially varying concentration of surface active molecules Marangoni flows effect the film drainage \cite{Karakashev2007}. Once the liquid film is below a critical thickness van-der-Waals forces lead to the rupture of the shared film \cite{Horn2011}.
Real liquid films contain impurities such as electrolytes, which affect the coalescence of bubbles \cite{Henry2008,Yaminsky2010}, too.

Besides the liquid properties, also the speed under which the bubbles approach each other affects the merging process \cite{Chan2011,Yaminsky2010, Ozan2019}. This can lead to a pimple, wimple, dimple, or ripple deformation of the liquid films \cite{Chan2011, Ozan2019}. A pimple results from a very slow approaching speed and a negative disjoining pressure such that the film surfaces attract each other \cite{Valkoska1999, Chan2011}. A wimple shows a varying film thickness. The central region is the thinnest and thickens radially \cite{Clasohm2005,Chan2011}. Ripples are interfaces called where a gap between the bubbles or a bubble and a substrate varies in thickness with multiple maxima and minima.
Soap bubbles, which merge with a flat film show a cascade of partial coalescence. This means a smaller bubble remains after the contact between bubble and film \cite{Pucci2015}.

Here, we report about the merging of two soap bubbles at low approach velocity. Then a dimple forms, its shape just before coalescence is revealed through interferometry. Once the two films connect on a point-like liquid bridge, surface tension accelerates a rim connecting the two films radially outward. This process is studied with high-speed photography and an instability observed. While surfactants are important to stabilize the bubbles, we also demonstrate that they play a crucial part in the hydrodynamic merging process.

\begin{figure}
\includegraphics[width=1\columnwidth]{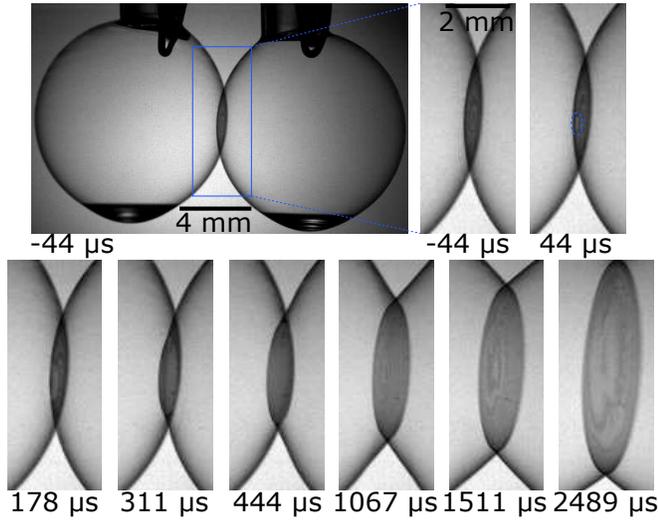}
\caption{Inclined side view of two merging soap bubbles. In the magnified view at $t=-44\,\upmu$s fringes can be observed in the area, where the bubbles touch each other. Time $t=0$ corresponds to the start of the liquid bridge formation. The dashed circle at $t=44\,\upmu$s surrounds the the area where the merging started. After coalescence the two bubbles share one film.}
\label{fig:overview}
\end{figure}

The soap bubbles are created by dipping tip of the syringes (\textit{Soft-Ject Insulin syringes with Luer connection, Henke-Sass, Wolf GmbH Germany}) into a soap solution and inflating the flat film formed by pressing on the plunger. A bubble attached to the syringe tip is inflated to about $12\,$mm in diameter. Once two soap bubbles are created this way they are brought together. One of the bubbles is stationary and the second bubble is moved slowly towards the first with a translation stage (\textit{M1 micromanipulator, Helmut Saur Laborbedarf, Germany}) at an approach velocity of $\approx 1\,$cm/s. During this approach the bubbles suddenly coalesce by forming a single soap film connecting the two bubbles (cf. supplementary material for a sketch of the experimental setup and Fig. \ref{fig:overview}).
The dynamics of coalescence is observed with a high-speed camera (Photron AX200) at 22,500 or 67,500 frames/s (exposure time 1/900,000\,s) and illuminated with either a white light source (Sugar Cube Ultra, \textit{USHIO AMERICA, INC.}, USA) or a coherent light source (\textit{CW532-04 Series, Roithner Lasertechnik GmbH, Austria}, CW laser, wavelength $\lambda = 532\,$nm, intensity $\approx\,2\,$mW/cm$^2$).
To avoid electric charging of the two bubbles during the inflation at nozzle of the plastic syringe \cite{Choi2013} we connect both wetted tips of the syringes with a copper wire that is held on a fixed electric potential.

The bubbles are made of an aqueous solution of the anionic surfactant sodium dodecyl sulfate (\textit{Sigma-Aldrich, BioXtra $\ge$99.0\% (GC)}, critical micelle concentration (cmc) 7-10\,mM at 20-25\,$^\circ$C). Two concentrations are used, namely 5 and 10\,mM and their coefficient of surface tension has been measured with the capillary rise method to be $\sigma=25\pm2$\,mN/m and $\sigma=27\pm2$\,mN/m for a 5 and 10\,mM  SDS solution, respectively. We measured their kinematic viscosity using an Ubbelohde viscometer, from which we calculated the dynamic viscosity, which is $1.01\,\pm\,0.02\,$mPa\,s for both solutions.

Figure \ref{fig:overview} is composed of selected snapshots from a high-speed imaging sequence showing the process of two bubbles merging. At the top of the first frame the syringe tips and the copper wire are visible. The bottom of this frame shows a darker area that is caused by liquid draining due to gravity. The first snapshot ($t=-44\,\upmu$s) is taken shortly before the bridging occurs. A weak fringe pattern is visible just before the two bubbles merge. The dashed circle at $t=44\,\upmu$s indicates the point where the liquid bridge has connected the two films. Between $t=44\,\upmu$s and $t=2489\,\upmu$s the two bubbles merge. A compound bubble is formed through the sharing of a liquid film. The merged film will grow until the angle between the two bubbles becomes 120$\,^{\circ}$. The concentric structure after $t=1067\,\upmu$s occurs due to thickness differences in the film. 

\begin{figure}
\centering
\includegraphics[width=1\columnwidth]{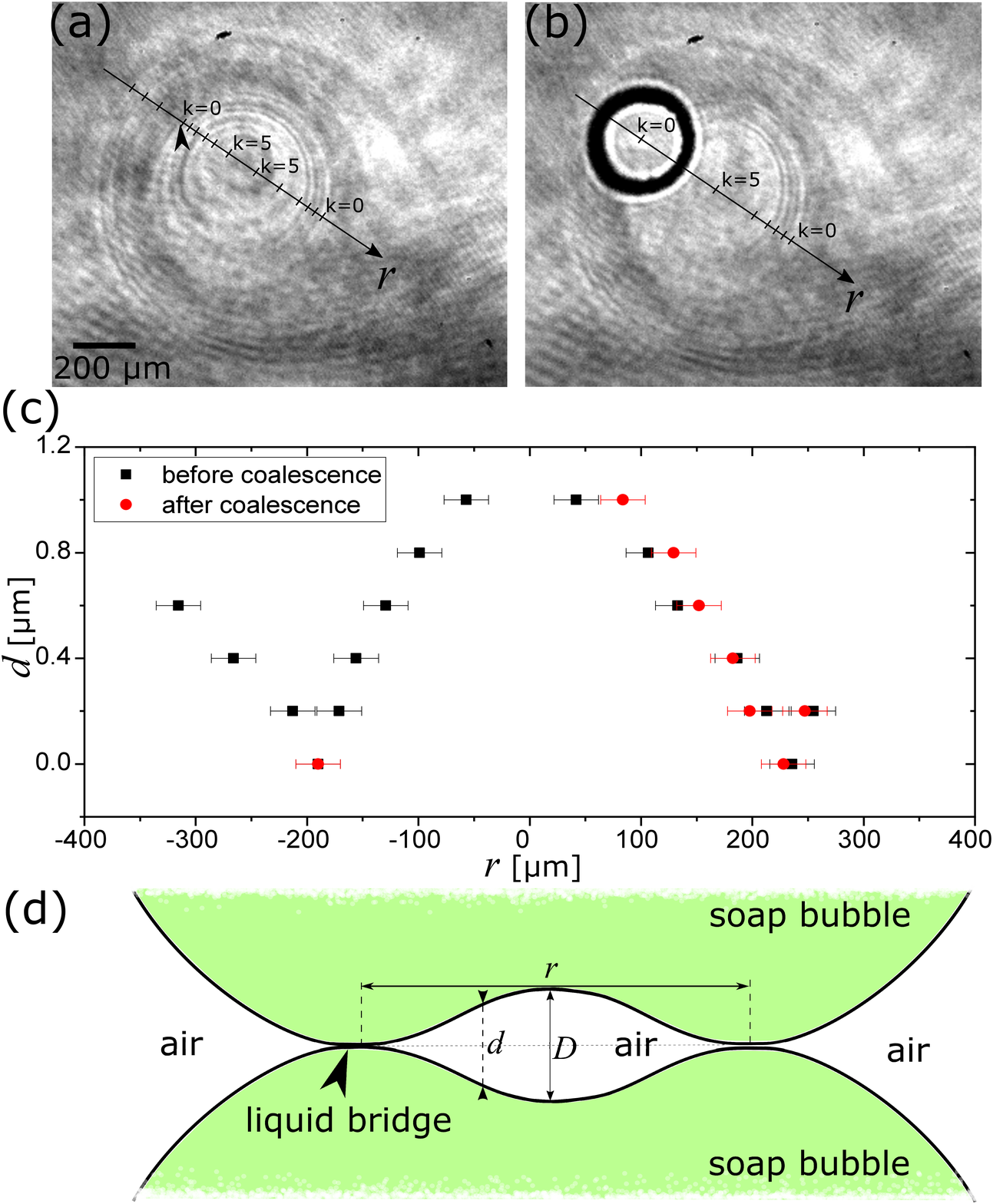}
\caption{(a) Interference fringes in the soap film preceding the coalescence and (b) in the frame after the coalescence (frame interval $\Delta t=44\,\upmu$s).(c) From the fringes the spatial varying height of the air film can be calculated and is shown before (black squares) and after coalescence (red circles). (d) A sketch of the overall shape of the dimple. Note that the vertical scale is strongly stretched.}
\label{fig:fringes}
\end{figure}
We are concerned with the initial process of film merging. For this we first want to understand the shape of the air gap separating the soap films. The interference fringes forming prior to the coalescence event are observed in Figure \ref{fig:fringes}a with an illumination of a continuous wave laser at $\lambda=532\,$nm wavelength. We observe in a magnified view the two films with a spatial resolution of $22\,\upmu$m/pixel. Unfortunately, the quality of the interference pattern is affected by the imaging and illumination through two soap films. Nevertheless, we observe concentric and mildly distorted circles where the distances between two fringes is decreasing towards the center, which is consistent with the presence of entrapped air between the two soap films. Figure \ref{fig:fringes}b just after coalescence demonstrates that the two films bridge between the crests of the dimple which is the shortest distance between the two films. Since the measurements are performed in transmission, the shortest distance has a bright fringe. That location is indicated in Fig. \ref{fig:fringes}a with an arrow and is the zeroth order fringe ($k=0$). The distance between bright fringes of $\lambda/2n$ ($n$ is the refractive index of the soap solution: $n=1.33$) allows converting the line drawn in Figs. \ref{fig:fringes}a and \ref{fig:fringes}b into a height map of the air gap before and just after coalescence, see Fig. \ref{fig:fringes}c. The position of the fringes are determined by taking the mean gray values along several lines through the fringe pattern. The maximum thickness of the air film is $1.0 \pm 0.1\,\upmu$m. Repeating the experiment 50 times we find dimple heights between $0.4\,\upmu$m and $1.0\,\upmu$m at the moment of coalescence. The overall shape of the gap between the two films is sketched in Fig. \ref{fig:fringes}d. Please note that the vertical axis is strongly magnified.

\begin{figure*}
\includegraphics[width=2\columnwidth]{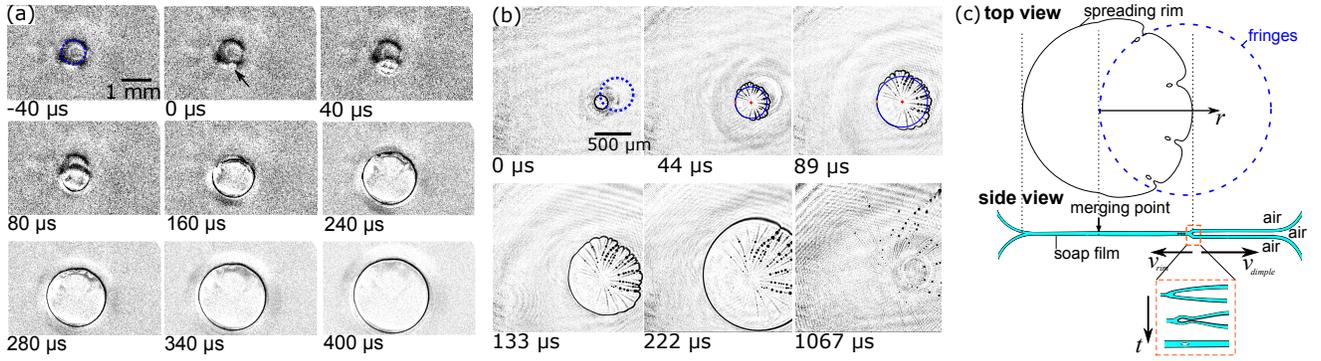}
\caption{Spreading of the merged films during the coalescence of two soap bubbles (a) with a concentration of [SDS]=5\,mM and (b) with [SDS]$=$10\,mM. The dark circle (blue dotted) in the first snapshot represents the fringe pattern. The films merge at $t=0$ in (a) at the location indicated by the arrow and in (b) at the lower left side of the fringes. In (b) at $t=44$ and 89\,$\upmu$s the smooth rim and the liquid bridge indicated with a blue circle and a red cross, respectively. The modulated rim that connects the two films propagates fastest through the dimpled region.
(c) Sketch of the spreading rim from top and in side view. The rim propagates faster within the regime of the dimple (dashed blue circle) due to the higher curvature. In the side view the region which is already merged consists only of a singe film, whereas still two separate films exist ahead of the rim. }
\label{fig:vgl_SDS}
\end{figure*}

We now discuss the growth of the merged films starting from the location where the two films are bridged. Figure \ref{fig:vgl_SDS}a shows a typical example of this merging dynamics. The dimpled region just before coalescence is indicated with a blue dotted circle and the arrow indicates the location of the liquid bridge. Bridging always happens on the crest of the dimpled region as shown in Fig. \ref{fig:fringes}. We see the rim expanding radially from the point of contact. Interestingly, that part of the rim which travels through the dimpled region has a distorted and fuzzy front, while the part that travels outside is smooth and circular. Figure \ref{fig:vgl_SDS}b shows a similar case with the only difference that the concentration of surfactant was doubled to [SDS]$=$10\,mM. The liquid bridge forms on the lower left part of the dimpled region, again on a crest and the rim spreads circularly. In contrast the rim traveling through the dimpled region reveals an instability or modulation with a length scale of $44\,\upmu$m at $t=89\,\upmu$s after bridging. Small dark structures pinch-off from the modulated rim leaving behind radial streaks. These streaks of round objects emanate from the slower parts of the modulated rim. Even after the rim has left the field of view at $t=1067\,\upmu$s in Fig. \ref{fig:vgl_SDS}b, radial pointing structures remain on the merged film. Figure \ref{fig:vgl_SDS}c sketches the process of film merging within the dimple (right) and outside the dimple (left). The part of the rim with a radial modulation has a non-uniform radial velocity. As a consequence the two films cannot merge simultaneously at a distance $r$ from the liquid bridge. Instead pockets of gas become entrapped during their merging. This is consisted with the observation that the round objects in Fig. \ref{fig:vgl_SDS}b are formed between the slowest traveling parts and the fastest parts of the rim. Connecting these clues we suggest that the objects are microscopic bubbles entrapped by the non-homogeneous merging of film. A close-up of the structured process is depicted in Fig. \ref{fig:vgl_SDS}b at $t=44$ and 89\,$\upmu$s where the rim traveling to the left is marked with a blue circle centered around the location of the liquid bridge (red cross). 
Comparing the radius of the circle and the radius of the modulated rim, it is clearly visible that the velocity in the dimpled regime is higher, i.\,e. 3.0$\pm$0.2\,m/s, whereas the velocity of the smooth rim is 2.5$\pm0.2\,$m/s. The difference in the velocity can be attributed to the varying curvature of the of the soap film inside and outside the dimple. Outside the two free soap films separate more rapidly than within the dimple. 

\begin{figure}
\includegraphics[width=1\columnwidth]{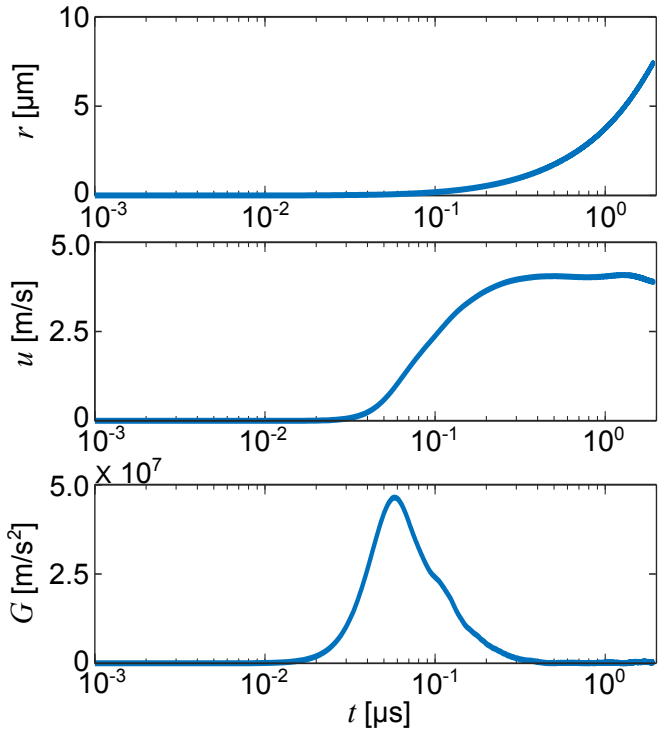}
\caption{Numerical simulations of the development of the rim radius over time (top row) and reaches a velocity of $4.0\,$m/s (center row). The rim accelerates within 0.5$\,\upmu$s to a velocity of 4.0\,m/s, i.\,e. that results in an acceleration of $5\cdot10^7$\,m/s$^2$ (bottom row). Dimple height: 1.6\,$\upmu$m, film thickness: 5.5\,$\upmu$m.}
\label{fig:simulations}
\end{figure}

We now address the mechanism destabilizing the rim's circular shape. For this we estimate the acceleration $G$ of rim due to surface tension $\sigma$. Ignoring viscosity we balance inertia with the pressure gradient from surface tension, i.\,e.
\begin{align}
G= \frac{\text{D}u}{\text{D}t}=-\frac {1}{\rho_H}\nabla p \approx - \frac {2 \sigma}{\rho_H D^2},
\label{eq:acceleration}
\end{align}
where $u$ is the velocity of the fluid particle in the rim, $\rho_H$ is the density of the liquid and $\nabla p$ the pressure gradient. The latter we estimate with the Laplace pressure in the cylindrical rim of cross-section $D$ and radius of curvature of $\approx D/2$. Inserting suitable values for $\sigma=0.025$\,N/m, $\rho=10^{3}$\,kg/m$^3$ and $D=1.6\cdot10^{-6}$\,m we obtain a high values for the acceleration with $G=2\cdot10^{7}$\,m/s$^2$. While viscosity counteracts this acceleration we nevertheless expect this acceleration to act until viscosity has diffused. This picture has been confirmed with Volume of Fluid simulations in axisymmetry (cf. supplementary material for details to the simulations). Figure \ref{fig:simulations} depicts a simulation result for the parameters similar to the experiment and accounting for viscous and compressible effects (solver Interfoam from the OpenFOAM framework). Within $0.5\,\upmu$s the rim accelerates from 0 to a velocity of about $4.0\,$m/s. 

Now we argue that this extreme acceleration of the rim from the liquid into the gas phase destabilizes the rim due to the Rayleigh-Taylor instability \cite{Zhou2017I,Zhou2017II}. The most amplified wavelength is $\sqrt{3} \lambda_c$ with $\lambda_c = \sqrt{\sigma/(G \rho_H)}$, see Ref. \cite{Sharp1984}. Typical experimental values lie between 44\,$\upmu$m (as in Fig. \ref{fig:vgl_SDS}b at $t=89\,\upmu$s) and 150\,$\upmu$m.
Since the rim expands and hence the length-scale of the instability, the most unstable mode (i.\,e. how many lobes the rim has as a result of the instability) is taken as a quantitative measure. We obtain values for the unstable mode between 12 (as in Fig. \ref{fig:vgl_SDS}a) and 44. 
Perturbations below a critical wavelength $\lambda_c$ are stabilized by surface tension. The critical wavelength can be calculated as \cite{Sharp1984}
\begin{align}
\lambda_c =  \sqrt{\sigma/G (\rho_H-\rho_L)},
\label{eq:lambda_c}
\end{align}
where $\rho_H$ and $\rho_L$ are the densities of the heavier (water) and lighter (air) fluids, respectively. 
Since $\rho_H \gg \rho_L$, we can neglect $\rho_L$ in equation \eqref{eq:lambda_c}. 

The acceleration $G$ governs the dominant wavelength at the rim, which strongly depends on the dimple height. With an acceleration of $G=2\cdot10^{7}$\,m/s$^2$ $\lambda_c$ is $\approx$ 0.7\,$\upmu$m.

The instability already occurs during bridging, and thus a diameter of the liquid bridge of 7\,$\upmu$m with the initial wavelength of $\lambda_c/\sqrt3=0.4\,\upmu$m would lead to a wavelength of 44\,$\upmu$m at a circumference of the rim of 2.4\,mm. Thus, $\lambda_c\approx 0.7\,\upmu$m seems an appropriate value. 

The  Rayleigh-Taylor instability leads to a pearling of tiny soap bubbles from the rim if the indentation is sufficiently large. The indentation leads locally to a higher curvature of the rim and thus a higher propagation velocity of the highly curved parts. The curvature decreases due to the reunion of the water columns and thus the velocity decreases. The rim becomes smooth after the propagation through the dimpled area and remains circular. 

Upon the approach of two merging bubbles a dimple is formed between them. This is a general phenomenon when bubbles or droplets approach either each other \cite{Klaseboer2000,Manica2008,Liu2018} or a substrate \cite{Fisher1991,Veen2012,Hendrix2012}.
The computed height of the dimple calculated using the Bragg equation agrees very well with the height in other experiments \cite{Veen2012,Hendrix2012}. The distance between the bubbles is shortest at the rim of the dimple and the film thickness at this point decreases, such that attractive van-der-Waals forces become important and lead to a rapid decrease of the film thickness \cite{Manica2007,Manica2008}. The point where the coalescence of the bubbles occurs is randomly distributed over the rim of the dimple. 

The applied potential on the bubbles ensures reproducible results, however, a simple connection between the bubbles (i.\,e. a wire to exchange electric charges) yields the same results. In experiments without this potential, a deformation of the rim is only observed by chance. We speculate that the bubbles are charged by inflating them. A similar effect was investigated by Choi et al. \cite{Choi2013}, who observed that droplets become charged during conventional pipetting.

The closure of the distorted rim depends strongly on the SDS concentration: with increasing amount of SDS more bubble pinch-off events are seen. However, when reaching the critical micelle concentration, a further increase of the bubble formation does not occur. The detachment of tiny soap bubbles from the rim is only possible, because more surface active molecules are available than necessary for the current surface. So the excess molecules can be used to create and stabilize a new surface in a much shorter time, which leads to the pearling of the droplets. The pearling is more pronounced directly after the formation of the liquid bridge than further away from the bridging point, since the distance of the films increases with distance from the bridging point and thus the closure dynamics slows down. 

The measured wavelength of the indentation agrees with an impulsively acting Rayleigh-Taylor instability and growth by radial expansion. The high acceleration of the liquid within the first 50\,ns after the coalescence, which is confirmed by simulations, induces the instability. Since the instability already occurs during bridging the instability timescale is consistent with the claim that the acceleration of the liquid is causing the instability \cite{Richtmyer1960,Ding2019,Zhou2019}. 

Structures similar to those described in the current work were already found in similar systems at the receding rim of bursting fluid films \cite{Reyssat2006, Lhuissier2009} or during the impact of a droplet on a fluid surface \cite{Thoroddsen2012}. Another work was provided by Thoraval et al., who investigated an impacting droplet on a liquid layer and found a formation of bubble rings in the impact region \cite{Thoraval2013}.
Despite these structures look similar to those described in the current work, the mechanism of their formation differs.

In this work the coalescence of soap bubbles is studied. Prior to their coalescence the bubbles form a dimple and entrap a tiny volume of air. Upon merging, the rim of the spreading film is accelerated for a brief moment, simulations predict less than $1\,\upmu$s. During that time a Rayleigh-Taylor instability sets in resulting in an instability of the rim front. The velocity of the rim is higher in the area of the dimple, due to a higher curvature in that regime and hence the velocity of the rim is faster. Depending on the surfactant concentration the entrapment of gas pockets is possible with increasing SDS concentration. However above the cmc no further effect of the SDS concentration on the instability of the rim is observed.

See supplementary material for a sketch of the experimental setup, details and snapshots of the simulations. 

The Deutsche Forschungsgemeinschaft, DFG, is acknowledged for support within project DA 2108/1-1.


\providecommand{\noopsort}[1]{}\providecommand{\singleletter}[1]{#1}%

\end{document}